\documentclass[aps,prd,showpacs,twocolumn,superscriptaddress,preprintnumbers]{revtex4-1} 

\makeatletter
\def\p@subsection{}
\makeatother

\usepackage{color,graphicx,amsmath,amssymb,lipsum}
\usepackage[colorlinks=true,citecolor=blue,linkcolor=blue,urlcolor=blue, backref=false,pdfborder={0 0 0}]{hyperref}

\usepackage[normalem]{ulem}
\usepackage[utf8]{inputenc} 
\usepackage{mathtools}
\usepackage{tabularx} 
\usepackage{tabu}
\usepackage{multirow}
\usepackage[table]{xcolor}
\usepackage{colortbl}
\usepackage{caption}
\usepackage{tikz}
\usepackage{bbm}
\usetikzlibrary{snakes}
\usetikzlibrary{decorations}
\usetikzlibrary{trees}
\usetikzlibrary{decorations.pathmorphing}
\usetikzlibrary{decorations.markings}
\usetikzlibrary{external}
\usetikzlibrary{intersections}
\usetikzlibrary{shapes,arrows}
\usetikzlibrary{arrows.meta}
\usetikzlibrary{calc}
\usetikzlibrary{shapes.misc}
\usetikzlibrary{decorations.text}
\usetikzlibrary{backgrounds}
\usetikzlibrary{fadings}
\usetikzlibrary{tikzmark,calc,arrows,shapes,decorations.pathreplacing}
\tikzset{
        cross/.style={cross out, draw=black, minimum size=2*(#1-\pgflinewidth), inner sep=0pt, outer sep=0pt},
	branchCut/.style={postaction={decorate},
		snake=zigzag,
		decoration = {snake=zigzag,segment length = 2mm, amplitude = 2mm}	
    }}

\usepackage{float}

\usepackage{bm}

\include{figures}

\usepackage{ulem}
\usepackage{diagbox}

\usepackage{mathrsfs}

\newcommand{\be}{\begin{equation}}
\newcommand{\ee}{\end{equation}}
\newcommand{\beqa}{\begin{eqnarray}}
\newcommand{\eeqa}{\end{eqnarray}}

\newcommand{\bseq}{\begin{subequations}}
\newcommand{\eseq}{\end{subequations}}

\renewcommand{\ln}{\mathop{\rm ln}\nolimits}

\renewcommand{\Im}{\mathop{\rm Im}\nolimits}

\definecolor{cornellRed}{HTML}{B31B1B}
\definecolor{cornellBlue}{HTML}{0068AC}
\definecolor{cornellGreen}{HTML}{6EB43F}
\definecolor{purple}{HTML}{66023C}

\def\gsim{\raise0.3ex\hbox{$\;>$\kern-0.75em\raise-1.1ex\hbox{$\sim\;$}}}
\def\lsim{\raise0.3ex\hbox{$\;<$\kern-0.75em\raise-1.1ex\hbox{$\sim\;$}}}

\def\beqn#1{\begin{equation}\label{#1}}
\def\eeqn{\end{equation}}

\def\beqa#1{\begin{eqnarray}\label{#1}}
\def\eeqa{\end{eqnarray}}

\def\Z2{$\mathcal{Z_2}$}

\newcommand {\ignore}[1]{}

\immediate\write18{texcount -inc -total -q -sum universality.tex > wordcount.txt}

\begin{document}

\preprint{MIT-CTP/5863}

\title{Resummation of Universal 
Tails in Gravitational Waveforms
}

\author{Mikhail M. Ivanov}
\email{ivanov99@mit.edu}
\affiliation{Center for Theoretical Physics, Massachusetts Institute of Technology, 
Cambridge, MA 02139, USA}
\affiliation{The NSF AI Institute for Artificial Intelligence and Fundamental Interactions, Cambridge, MA 02139, USA}
\author{Yue-Zhou Li}
\email{liyuezhou@princeton.edu}
\affiliation{Department of Physics, Princeton University, Princeton, NJ 08540, USA}
\author{Julio Parra-Martinez}
\email{julio@ihes.fr}
\affiliation{Institut des Hautes Études Scientifiques, 91440 Bures-sur-Yvette, France}
\author{Zihan Zhou}
\email{zihanz@princeton.edu}
\affiliation{Department of Physics, Princeton University, Princeton, NJ 08540, USA}

\begin{abstract} 

We present a formula for the universal anomalous scaling of the multipole moments of a generic gravitating source in classical general relativity.
We derive this formula 
in two independent ways using effective field theory methods.
First, we 
use the absorption of low frequency 
gravitational waves 
by a black hole to 
identify the total multipole 
scaling dimension as the renormalized angular momentum of black hole perturbation theory. 
More generally, we show that the anomalous dimension
is determined 
by phase shifts 
of gravitational waves elastically scattering off
generic source multipole moments,
which reproduces the renormalized angular momentum
in the particular case of black holes.
The effective field theory approach thus clarifies 
the role of the renormalized angular momentum in the multipole expansion.
The universality of the point-particle effective description of compact gravitating systems further allows us to extract the universal part of 
the 
anomalous dimension, which is the same for any object, including black holes, neutron stars, and binary systems. As an application, we propose a novel resummation of the universal short-distance logarithms (``tails'')
in the gravitational waveform of binary systems, 
which may improve the modeling of signals from current and future gravitational wave experiments.
\end{abstract}

\maketitle

\textit{Introduction and Executive Summary.}--
The detection of 
gravitational waves (GWs)
from inspiraling black holes (BHs)
by the LIGO/Virgo/KAGRA experiment has brought about the era of 
precision strong-gravity science \cite{LIGOScientific:2014pky,VIRGO:2014yos,LIGOScientific:2016aoc,LIGOScientific:2018mvr,LIGOScientific:2020ibl,LIGOScientific:2021usb,LIGOScientific:2021djp,KAGRA:2020agh}. 
The two-body problem cannot be solved
exactly in full general relativity (GR), so
a number of approximate techniques have been 
utilized for precision calculations
of gravitational waveforms to model
the observed GW signals \cite{Blanchet:1985sp,Blanchet:1986dk,Rothstein:2014sra,Goldberger:2022rqf,Goldberger:2022ebt,Porto:2016pyg,Kalin:2020mvi,Mogull:2020sak,Cheung:2018wkq,Kosower:2018adc,Bern:2019crd,Bern:2020buy,Buonanno:2022pgc,Cheung:2023lnj,Kosmopoulos:2023bwc,Bjerrum-Bohr:2022blt,Brandhuber:2023hhy,Herderschee:2023fxh,Elkhidir:2023dco,Georgoudis:2023lgf,Caron-Huot:2023vxl,Buonanno:1998gg,Buonanno:2000ef}. One of such techniques
is gravitational effective field theory
of inspiraling binaries (EFT), which applies quantum field theory tools to this problem \cite{Goldberger:2004jt,
Goldberger:2005cd,
Porto:2007qi,
Porto:2016pyg,
Goldberger:2020fot,Goldberger:2022rqf,Goldberger:2022ebt}. The EFT approach 
makes the relevant degrees 
of freedom and underlying
symmetries of the problem manifest, and allows for a systematic
treatment of large- and small-scale 
divergences that appear
in the perturbative description of the 
GW emission. 
The effective action describing the GWs emitted by 
the binary is given by the Einstein-Hilbert term together with the wordline effective action
\be 
\label{eq:action}-\int d\tau\left[\mathcal{E}+\frac{1}{2}\omega_{ij}J^{ij}+\frac{1}{2}Q^{E}_{ij}E^{ij}+\frac{1}{2}Q^{B}_{ij}B^{ij}+...\right]\,,
\ee 
where $\tau$ is proper time, $\mathcal{E}$, $J^{ij}$
are the total conserved energy and the angular momentum of the binary, $\omega_{ij}$ is its angular velocity,
$Q^{E/B}_{ij}$ are its gravitational 
electric and magnetic 
quadrupole moments, 
while $E^{ij}, B^{ij}$ are the electric and magnetic 
parts of the Weyl curvature tensor describing the
emitted radiation. The dots stand for the coupling to higher
 multipole moments as well as tidal operators which depend on higher-powers of the curvature.
 We note that the multipoles generically
include spin-induced contributions \cite{Poisson:1997ha,Porto:2005ac,Marsat:2014xea, Levi:2014gsa,Levi:2015msa, Krishnendu:2017shb, Krishnendu:2018nqa,Chia:2020psj, Chia:2022rwc, Lyu:2023zxv}.

The above action 
simply encodes the fact that the binary seen from 
large distances 
can be approximated as a point
source with given energy, angular momentum
and a collection of multipole moments attached to it. 
This approximation is adequate for GWs  emitted
during the inspiraling phase as their 
wavelength $\lambda$ is much greater than the 
size of the binary $r$, or have   
frequency  $\omega \ll r^{-1}$.

Relativistic corrections to 
the physical waveforms 
arising from the non-linear structure 
of GR are interpreted as classical 
``loop corrections'' in EFT.
Among these corrections, particularly interesting ones are logarithmic terms that arise from small-scale
``ultraviolet'' (UV) parts of EFT
integrals,
which are referred to as (dissipative) ``tails'' \footnote{These are to be distinguished from conservative tail logarithms that arise in the binding energy and angular momentum of the binary \cite{Goldberger:2012kf,Blanchet:2019rjs}, e.g., in Eq.~\eqref{eq:EJcons}.}.
Such tails, when combined with a near-zone description of the binary (valid for frequencies $\omega \sim r^{-1}$) yield finite logarithmic contributions to the waveform \cite{Blanchet:1985sp,Blanchet:1986dk}. However, the binary's near zone is not described by the EFT in Eq.~\eqref{eq:action}, and hence the tails manifest as UV divergences in the effective description. This is, in fact, not a ``bug'' but a feature of the EFT description, which allows us to interpret these divergences as the (classical) renormalization group running of the radiative multipoles \cite{Goldberger:2009qd}, thus enabling the resummation of the associated logarithms in the waveform or other observable quantities.

For simplicity, let us use the harmonic space 
counterparts to the quadrupole moments $Q^{E/B}_{2 m}$ (where $m$
is the azimuthal harmonic number)~\cite{Thorne:1980ru,Charalambous:2021mea,Glazer:2024eyi}, which satisfy 
the following renormalization
group (RG)
equation \cite{Goldberger:2009qd}:
\be 
\label{eq:RGQ}
\mu\frac{d}{d\mu} Q^{E/B}_{2 m} = \gamma_{{2 m}}^{E/B} Q_{2 m}^{E/B}\,,
\ee
where $\mu$ is the matching scale, and $\gamma_{2 m}$ is the multipole  
anomalous dimension, which is defined by this equation and captures the coefficient of the dissipative tail logarithms in the waveform.  Previous calculations found $\gamma^{E/B}_{2m}=-\frac{214}{105}(G\mathcal{E}\omega)^2$ at the lowest order~\cite{Blanchet:1997jj,Goldberger:2009qd},
where $G$ is Newton's constant.  The physical 
interpretation of this RG equation
is that the relativistic 
gravitational potential of the binary 
effectively 
adds up to the quadrupole moment,
leading to more radiation emitted. 
Separating the potential modes from the source quadrupole, i.e., ``integrating the potential modes out'',
by lowering $\mu$ (or, equivalently, by increasing the size of the near zone)
increases the effective quadrupole moment,
giving rise to its scale dependence
akin to the scale dependence
of the coupling constant in 
quantum chromodynamics.  The multipole of a generic gravitating system satisfy such 
RG equation, as it stems
from the interaction of the radiation with
relativistic potential fields sourced by the binding energy and angular momentum
in eq.~\eqref{eq:action},
which is fixed by the non-linear structure of GR. 

While some higher order results for the 
anomalous dimension of the quadrupole and higher multipole moments have been derived in
the literature~\cite{Almeida:2021jyt,Trestini:2023wwg, Edison:2023qvg,Edison:2024owb}, in this letter we
present 
\textit{an exact formula}
for the anomalous dimensions of all multipoles.
Namely, we show that the 
anomalous dimension 
of the multipoles with angular and azimuthal numbers
($\ell$,$m$)
is determined by the partial wave phase shift $\delta_{\ell m}$ 
of gravitational waves elastically 
scattering off the system (see e.g., Eq.~(17) in Ref.~\cite{Ivanov:2024sds}), 
\be 
\label{eq:main_anom}
    \gamma_{{\ell m}}^{E/B} = -\frac{1}{\pi} \Big(\delta_{\ell m}^{E/B} (\omega) + \delta^{E/B} _{\ell m}(-\omega) \Big)\,,
\ee
where $\delta^{E/B} _{\ell m}(-\omega)$
is the phase shift describing the time reversal of the same scattering process. 

The equivalence principle dictates that Eq.~\eqref{eq:action}
describes any
system interacting with the long-wavelength 
GWs. 
This implies that the universal part of 
the anomalous dimension in Eq.~\eqref{eq:main_anom}
can be extracted from any gravitational scattering process.
This is true for both emission
and absorption of gravitational
waves. The equivalence between the two
is akin to relations between
Einstein coefficients for  
absorption and stimulated emission in atomic physics \cite{Einstein:1917zz}. 
In particular, we can use a simple 
problem with a known result:
the Raman (i.e., inelastic)
scattering of long-wavelength GWs off a solitary BH to read off the universal part of the anomalous dimension.
The scattering amplitudes 
of this process can be computed \textit{exactly}
using black hole perturbation theory
(BHPT)~\cite{Teukolsky:1972my,Teukolsky:1973ha,Teukolsky:1974yv,Mano:1996gn,Mano:1996mf,Mano:1996vt,Sasaki:2003xr,Aminov:2020yma,Bonelli:2021uvf,Bonelli:2022ten,Bautista:2023sdf}. 
%
%
%

It is known from BHPT that 
the scale dependence of BH multipole
moments in the near zone 
is captured by the so-called
``renormalized angular momentum'' $\nu$~\cite{Mano:1996gn,Mano:1996mf,Mano:1996vt,Sasaki:2003xr},
which depends on the BH mass $M$, spin $\chi=S/(GM^2)$,
and the GW frequency $\omega$ (see e.g.~\cite{Chakrabarti:2013lua,Chia:2020yla,Charalambous:2021mea}). 
Our relation between the anomalous dimension and scattering phase shows that 
this quantity precisely determines the 
anomalous dimension of BH multipole moments, 
\be 
\label{eq:bh_ad}
    \gamma_{{\ell m}}^{\rm BH} = \nu(GM\omega, \chi) - \ell\,,
\ee
which can be computed 
analytically for generic $\ell$
to any given order in $GM\omega$. 
Below we provide an independent argument for this based directly on the absorptive scattering of gravitational waves by BHs.

Note that the multipole moments in Eq.~\eqref{eq:action}
depend on the system's spin. 
The rotating (Kerr) BHs
obey the ``no-hair'' theorem~\cite{Teukolsky:2014vca}, dictating that 
all of their multipole moments are uniquely fixed by their spin and mass. 
This is not true for a generic gravitating system, 
for which spin-induced multipole moments \cite{Levi:2015msa} and tidal effects (Love numbers) \cite{Damour:1982wm,Damour:2009vw, Damour:1998jk, Damour:2009va, Binnington:2009bb,  Goldberger:2004jt,Kol:2011vg,Hui:2020xxx,Charalambous:2021mea,Charalambous:2021kcz,Charalambous:2022rre,Charalambous:2023jgq,Hui:2021vcv,Ivanov:2024sds}
break the universality
starting at 
$\mathcal{O}(J^2)$ and $\mathcal{O}(G^{2\ell+1})$ respectively.
This means that the multipole anomalous dimension of a BH is actually universal
through $\mathcal{O}(J G^{2\ell+1})$. Such universal anomalous dimension, valid for a general system, is then obtained simply by expanding the BH anomalous dimension to the appropriate order and replacing $M\to {\cal E}$ and $\chi \to {\cal J} \equiv J/(G{\cal E}^2)$, which yields the universal
anomalous dimension
\be  
\gamma_{\ell m}^{\rm univ.} \!=\! \left[\gamma_{\ell m}^{\rm BH}(G\mathcal{E} \omega,0) \!+\! \partial_{\chi  }\gamma_{\ell m}^{\rm BH}(G\mathcal{E} \omega,0) {\cal J}\right]_{G^{2\ell+1}}\,,\label{eq: universal ano}
\ee 
where $[\cdots]_{G^n}$ denotes an expansion through order $n$.
The knowledge of this anomalous dimension allows for a
resummation of the universal
short-distance
logarithms (tails) by means
of the RG Eq.~\eqref{eq:RGQ}. 
We claim that this result holds true
for \textit{any} gravitating system: a BH, 
a neutron star, or an inspiraling binary system.

In the rest of this letter, we provide an explicit 
proof to the above statements and 
produce waveform predictions
that contain the resummation of 
the universal UV tails. We first connect the anomalous dimension of BH
 multipoles 
with the BHPT renormalized angular momentum using the GR results on the 
inelastic scattering of GW by 
BHs. 
After that we show 
that the renormalization of radiative multipoles
is fixed by the scattering phase shift in Eq.~\eqref{eq:main_anom}, and explain which terms in the BH anomalous dimension are universal. 
Finally, we 
present an analytic formula for the binary waveforms
that includes the resummed universal
tails.

\textit{Black hole anomalous dimension from the renormalized angular momentum.--}
Let us start by showing that the anomalous
dimension of a generic  
BH multipole
moment is given by the BHPT renormalized angular momentum. 
To that end we use the EFT action~\eqref{eq:action}
with the full collection of 
internal multipole moments.
These multipole moments describe the absorption
of waves by the BH horizon, 
producing inelasticity 
in the Raman scattering amplitude~\cite{Goldberger:2020fot,Ivanov:2022qqt,Ivanov:2024sds}. 
Let us focus now on the electric moment only.
The computation for the magnetic part is identical.
The inclusive absorption cross-section
in the EFT 
at the tree-level is given by
the sum of the on-shell single-graviton 
absorption amplitudes over different
internal excited states $X$ of the back hole~\cite{Goldberger:2020fot, Goldberger:2022ebt}:
\be 
\label{eq:abs_singl}
\begin{split}
\sigma_{\rm abs} & =  
\lim_{T \to \infty}
\sum_{X}\frac{|\mathcal{M}_{\rm abs}(M\to X)|^2}{2\omega T}
\\
&
=\sum_\ell \frac{\ell !\omega^{2\ell +1}}{\omega^2(2\ell+1)!!}\sum_{m=-\ell}^\ell
\langle Q_{\ell m}Q_{\ell m}\rangle(\omega)\,,
\end{split}
\ee 
where we have introduced the Fourier
image of the in-in Wightman correlator $\langle Q_{\ell m}Q_{\ell m}\rangle(\omega)\equiv \int dt e^{i\omega t}
\langle
Q_{\ell m}(t)Q_{\ell m}(0)
\rangle $
in the initial state of the black hole. We switch now
to the partial wave absorption rate
\be 
\Gamma_{\ell m}\Big|_{\rm tree} = \omega^{2\ell+1}\langle Q_{\ell m}Q_{\ell m}\rangle(\omega)\,.
\ee 
Since we are computing classical observables, 
we can equivalently replace the Wighman correlator
above with the imaginary part of the retarded two-point function, which describes
classical absorption, 
$\Gamma_{\ell m}=\omega^{2\ell+1}\Im G^R_{\ell m}(\omega)$. 
In this sense we will refer to 
$Q_{\ell m}$ above as ``absorptive''
multiple moments $Q^{\rm abs}$. 


At higher order in the EFT, the absorption is in general 
described by diagrams by multiple
insertions of the $\langle Q Q \rangle$
correlators dressed with the potential gravitons. Let us 
consider diagrams with a single insertion of $\langle Q Q \rangle$.
Eq.~\eqref{eq:abs_singl} implies 
that its gravitational dressing can be represented  
as radiative corrections to the single graviton
absorption diagram, which are the same as the emission diagrams.
For instance the 
radiative corrections at order $G^2$ are given by
\begin{widetext}
\begin{equation}
i \mathcal{M}_{\rm abs}^{G^2}(\omega) =
\begin{gathered}
\begin{tikzpicture}[line width=1,photon/.style={decorate, decoration={snake, amplitude=1pt, segment length=6pt}}]
    \draw[line width = 1, photon] (0,-0.5) -- (1,-1.5);
    \draw[line width = 1, dashed] (0,-1.0) -- (0.5,-1.0);
    \draw[line width = 1, dashed] (0,-1.7) -- (0.5,-1.0);
    \filldraw[fill=white, line width=1.2](0,-0.5) circle (0.15) node[left]{\small$Q^{\rm abs}\;$};
    \filldraw[fill=gray!5, line width=1.2](0,-1.0) circle (0.15) node {$\times$};
    \node[left] at (0,-1.0) {\small $M(S)$\;};
    \filldraw[fill=gray!5, line width=1.2](0,-1.7) circle (0.15) node {$\times$};
    \node[left] at (0,-1.7) {\small $M$\;};
    \end{tikzpicture}
\end{gathered}
+
\begin{gathered}
    \begin{tikzpicture}[line width=1,photon/.style={decorate, decoration={snake, amplitude=1pt, segment length=6pt}}]
    \draw[line width = 1, photon] (0,-0.5) -- (1,-1.5);
    \draw[line width = 1, dashed] (0,-1.2) -- (0.5,-1.0);
    \draw[line width = 1, dashed] (0,-1.7) -- (0.8,-1.3);
    \filldraw[fill=white, line width=1.2](0,-0.5) circle (0.15);
    \filldraw[fill=gray!5, line width=1.2](0,-1.2) circle (0.15) node {$\times$};
    \filldraw[fill=gray!5, line width=1.2](0,-1.7) circle (0.15) node {$\times$};     \end{tikzpicture}
\end{gathered}
~, \qquad 
i \mathcal{M}_{\rm emi}^{G^2}(\omega) = 
\begin{gathered}
\begin{tikzpicture}[line width=1,photon/.style={decorate, decoration={snake, amplitude=1pt, segment length=6pt}}]
    \draw[line width = 1, photon] (0,-0.5) -- (1,-1.5);
    \draw[line width = 1, dashed] (0,-1.0) -- (0.5,-1.0);
    \draw[line width = 1, dashed] (0,-1.7) -- (0.5,-1.0);
    \filldraw[fill=black, line width=1.2](0,-0.5) circle (0.15) node[left]{\small$Q^{\rm rad}\;$};
    \filldraw[fill=gray!5, line width=1.2](0,-1.0) circle (0.15) node {$\times$};
    \node[left] at (0,-1.0) {\small ${\cal E}(J)$\;};
    \filldraw[fill=gray!5, line width=1.2](0,-1.7) circle (0.15) node {$\times$};
    \node[left] at (0,-1.7) {\small $\cal E$\;};
    \end{tikzpicture}
\end{gathered}
+
\begin{gathered}
    \begin{tikzpicture}[line width=1,photon/.style={decorate, decoration={snake, amplitude=1pt, segment length=6pt}}]
    \draw[line width = 1, photon] (0,-0.5) -- (1,-1.5);
    \draw[line width = 1, dashed] (0,-1.2) -- (0.5,-1.0);
    \draw[line width = 1, dashed] (0,-1.7) -- (0.8,-1.3);
    \filldraw[fill=black, line width=1.2](0,-0.5) circle (0.15);
    \filldraw[fill=gray!5, line width=1.2](0,-1.2) circle (0.15) node {$\times$};
    \filldraw[fill=gray!5, line width=1.2](0,-1.7) circle (0.15) node {$\times$};     \end{tikzpicture}
\end{gathered}\,.
\label{fig:dress}
\end{equation}
\end{widetext}
The equivalence between
the radiative corrections 
to the BH 
absorption and 
the emission of gravitational 
waves by a binary is a simple
fact that follows
from the universality 
of the action~\eqref{eq:action}.
The sum of Feynman 
diagrams corresponding to
the single $\langle QQ \rangle$
insertion yields
\be 
\label{eq:eft_logs}
\Gamma_{\ell m}\Big|_{\rm tree}^{\rm dressed} = 
\Gamma_{\ell m}\Big|_{\rm tree} \left(1+\sum_{n=1} \epsilon^n \sum_{k}a_k \ln^{b_k}\left(
\frac{\omega}{\mu}\right)\right)\,,
\ee 
where $\epsilon=GM\omega$, while $a_k,b_k$
are numerical coefficients. For clarity, we 
omitted the unobservable IR logs, so that 
all the logs above 
stem from UV divergences. 
The coefficients $a_n$ above are 
in general divergent and they are to be renormalized via
a multiplicative wavefunction renormalization~\cite{Goldberger:2009qd}. 
The renormalized EFT expression can be compared with the BHPT  result~\cite{Mano:1996gn,Mano:1996mf,Mano:1996vt,Ivanov:2022qqt}, 
\be 
\label{eq:bhpt_abs_schem}
\Gamma_{\ell m}\Big|_{\rm BHPT} = 
(GM\omega)^{2\nu+1}\frac{\mathcal{A}_\omega}{|1+(GM\omega)^{2\nu+1} \mathcal{B}_\omega|}\,,
\ee 
where $\mathcal{A}_\omega, \mathcal{B}_\omega$ 
are 
power 
series in $\omega$, while $\nu$ is the frequency-dependent renormalized angular momentum,
\be 
\nu =\ell -\frac{2 \left(15 \ell ^2 (\ell +1)^2+13 \ell  (\ell +1)+24\right)}{(2
   \ell +1) \ell  (\ell +1)  (4 \ell  (\ell +1)-3)}\epsilon ^2 
+ \cdots\,,
\ee 
giving $\nu = 2 -\frac{214}{105}(GM\omega)^2 +\cdots$ for $\ell=2$.
The $\nu$ exponent is the only source of non-analyticity 
in eq.~\eqref{eq:bhpt_abs_schem}
that generates all the logs 
in the low-frequency expansion~\eqref{eq:eft_logs}.
Thus, $(GM\omega)^{2\nu+1} \mathcal{A}_\omega$
is identified as the radiatively corrected 
EFT multipole two point function, while the integer 
powers of $(GM\omega)^{2\nu+1}$ 
correspond to diagrams with multiple
insertions of the worldline multipole correlators.

Focusing on the single-multipole
term in eq.~\eqref{eq:bhpt_abs_schem}, factoring out the tree-level part, 
and formally inserting the matching scale $\mu$ we 
split the above formula
into the EFT (or far zone) and UV (or near zone) parts: 
\begin{align}
\label{eq: GR phase shift}
   & \Gamma_{\ell m}= (GM\omega)^{2\ell+1}(GM\omega)^{2(\nu-\ell)}\mathcal{A}_\omega\\
   &=\left(\frac{\omega}{\mu}\right)^{2(\nu-\ell)} 
   \!(GM\mu)^{2(\nu-\ell)} \omega^{2\ell+1}\text{Im}G^R_{\ell m}(\omega)(1\!+\!\sum c_n \epsilon^n) \nonumber\\
   &= \left(\frac{\omega}{\mu}\right)^{2(\nu-\ell)} 
 (1+\sum \tilde c_n \epsilon^n) \omega^{2\ell+1}  \langle Q^{\rm ren.}_{\ell m}Q^{\rm ren.}_{\ell m}\rangle(\omega,\mu)\,, \nonumber
\end{align}
where $(\omega/\mu)^{2(\nu-\ell)}$ represents the sum of UV tails 
from the logarithmically divergent loop integrals,
$\sum c_n\epsilon^n$ denotes the finite loop corrections,\footnote{Switching from $\sum c_n\epsilon^n$
to $\sum \tilde c_n\epsilon^n$ takes into account
that the renormalized multipoles absorb some 
scheme-dependent
finite
loop parts.} 
while $Q^{\rm ren.}_{\ell m}$ is the renormalized scale-dependent multipole,
\be 
Q^{\rm ren.}_{\ell m}(\omega;\mu) = \left(\frac{\mu}{\mu_0}\right)^{\nu -\ell}Q^{\rm ren.}_{\ell m}(\omega;\mu_0)\,,
\ee  
with the matching scale comparable with the inverse size of the BH, $\mu_0 \sim (GM)^{-1}$. The above implies
the following RG flow of the
absorptive
multipole moment operator, 
cf. Eqs.~\eqref{eq:RGQ},~\eqref{eq:bh_ad},
\be 
    \frac{d Q^{\rm ren.}_{\ell m} (\omega; \mu)}{d \log \mu} = (\nu -\ell) 
    Q^{\rm ren.}_{\ell m}(\omega; \mu)\,,
\ee 
which resums tail logarithms associated to UV divergences in diagrams such as those in Eq.~\eqref{fig:dress}.
There are additional logarithms that are produced when 
the series expansion of $\nu$ 
in Eq.~\eqref{eq:bhpt_abs_schem}
hits poles
in $\mathcal{B}_\omega$. These correspond to 
the non-universal pieces described by EFT diagrams 
that feature the insertion of
dynamical tidal 
Love number operators, see e.g.,~\cite{Ivanov:2024sds,Caron-Huot:2025tlq}. 
This issue manifests itself 
as poles for integer values of $\ell$
in the perturbative expansion of $\nu$ at starting order $G^{2\ell+3}$.

\textit{Renormalization of radiative multipoles from scattering.--}
Let us now give a more general argument
that will confirm the equivalence 
between the 
renormalized angular momentum
and the multipole anomalous dimensions.
The emission of 
gravitational waves in the far zone of a generic system is controlled by the radiative multipoles $Q_{\ell m}^{\rm rad}$ \cite{Goldberger:2004jt,Goldberger:2009qd,Goldberger:2022rqf}. For instance, the leading order is described by the Einstein quadrupole formula.
The universality of the worldline EFT suggests that the tail effects in the inspiral binary waveforms originate from 
short-distance (near-zone) corrections to 
the radiative multipoles~\cite{Goldberger:2009qd}. 
A more general formula for the anomalous dimension of multipoles in terms of scattering phase shifts is provided by Eq.~\eqref{eq:main_anom}. We now derive this formula.

Let us consider how gravitational waves emitted from the radiative multipoles travel through the gravitational background of the binary out to infinity.
This process is described by the following local worldline EFT operators
$\mathbf{O}^{E}_{\ell m}(\tau) = Q_{\ell m}^{\rm E, rad} E_{\ell m}$ and $\mathbf{O}^{B}_{\ell m}(\tau) = Q_{\ell m}^{\rm B, rad} B_{\ell m}$. 
It is useful to consider the symmetric (Keldysh) correlator,
which is manifestly time-reversal invariant
\begin{equation}
    G^P_S(\omega) = \frac{1}{2} \langle 
    \{\mathbf{O}^P_{\ell m}(-\omega),\mathbf{O}^P_{\ell m}(\omega)\}|
    \rangle \,\label{eq: GS}\,,
\end{equation}
with parity $P=\pm 1$ for $E/B$.
This correlator captures the intrinsic fluctuations of compact objects in a gravitational-wave background, across time and energy scales.
Since the classical tidal fields do not experience classical RG running, the dilatation operator, $D \equiv \omega \partial_\omega =  - \mu \partial_\mu$, acts trivially on the gravitational field, which implies that this correlator satisfies the RG evolution equation dictated by the multipole moments
\begin{equation}
\frac{dG^P_S(\omega;\mu)}{d\log \mu} = 2 \gamma^P_{\ell m}(\omega)\,G^P_S(\omega;\mu) ~.
\end{equation}

To relate the anomalous dimension for the radiative multipoles to scattering, we follow the ideas of \cite{Caron-Huot:2016cwu} and study the analytic dependence of this correlator as a function of the frequency, $\omega$. In particular we consider the 
analytic continuation to negative frequencies, 
\begin{equation}
    \omega \to e^{i\pi} \omega\,. 
\end{equation}
On the one hand, by making use of the dilatation operator,  such analytic continuation extracts the anomalous dimension as a phase
\begin{equation}
    G^P_S(e^{i\pi}\omega) = e^{i \pi  D} G^P_S(\omega) = e^{ i \pi 2\gamma_{\ell m}^P} G^P_S(\omega)\,.
\end{equation}
On the other hand, the analytically continued correlator is simply related to its complex conjugation
\begin{equation}
\begin{aligned}
\label{eq:gstargamma}
   G^P_S(e^{i\pi}\omega) & = G^*_S(\omega) 
    = \frac{1}{2} \langle 
    \{\mathbf{O}^{P \dagger}_{\ell m}(-\omega),\mathbf{O}^{P \dagger}_{\ell m}(\omega)\}
    \rangle \,.
\end{aligned}
\end{equation}
Considering the insertion of operators $\mathbf{O}^P$ as a perturbation to the $S$-matrix describing the scattering of gravitational waves by the system,  $\delta S = i\mathbf{O}^{P\dagger} $, unitarity then implies
\begin{equation}
\mathbf{O}^{P\dagger}_{\ell m} = S^\dagger \mathbf{O}_{\ell m}^P S^\dagger \,.
\end{equation}
Inserting this relation into Eq.~\eqref{eq:gstargamma} 
and using the partial wave basis, we find 
\begin{equation}
\label{eq:gstardelta}
    G^P_S(e^{i\pi}\omega) = e^{-2i (\delta_{\ell m}^{P}(\omega) + \delta_{\ell m}^{P}(-\omega))} G_S(\omega)  ~,
\end{equation}
where $\delta_{\ell m}^{P}(-\omega)$ is the analytic continuation of the phase shift performed with fixed ${\cal J}$, that is, the time-reversed phase shift.  
Comparing Eqs.~\eqref{eq:gstargamma} and~\eqref{eq:gstardelta} we find that the anomalous dimension is directly related to the phase shift, as advanced in Eq.~\eqref{eq:main_anom}.
This is an exact relation for the anomalous dimension of radiative multipoles, valid for a generic system.

For the specific case of BHs, using the known formulae for Raman scattering from
BHPT
(see e.g., 
Eq.~(4.3) of Ref.~\cite{Bautista:2023sdf} and Eq.(3.13) in \cite{Saketh:2023bul}), we recover
the claimed result for the BH anomalous dimension in Eq.~\eqref{eq:bh_ad}
for generic $\ell$.


\textit{Universal anomalous dimension from black holes.--}
Let us now discuss 
to what extent the exact result for the anomalous dimension of BH mutipole moments in Eq.~\eqref{eq:bh_ad} applies 
to generic systems. First of all, while the 
nonlinear interactions with the 
energy term in the action~\eqref{eq:action}
are the same for any system,
the angular-momentum (spin)
dependent terms beyond the 
linear one are specific
to a gravitating source.
Furthermore, it is important to note that the scattering phase shift, as computed in the EFT, receives contributions from non-universal tidal Love numbers starting at ${\cal O}(G^{2\ell+1})$. The leading contribution from these is odd in the frequency and hence cancels in Eq.~\eqref{eq:main_anom}. However, starting at the next order, diagrams containing the tidal operators will generate non-universal corrections to the anomalous dimension. The situation is even worse starting at ${\cal O}(G^{2\ell+3})$ where UV divergences in the far-zone phase shift appear, requiring the introduction of (running) dynamical Love numbers~\cite{Goldberger:2020fot,Saketh:2023bul,Ivanov:2024sds,Caron-Huot:2025tlq}.

Hence, the universal part
of the anomalous dimension
can be extracted from that of 
BH via a formal Taylor 
expansion in spin and $G$:
\begin{align}
    \gamma^{\rm BH}_{\ell m}(G M \omega,\chi) 
    & = \left[\gamma^{\rm BH}_{\ell m}(G M \omega,0) + \partial_\chi\gamma^{\rm BH}_{\ell m}(G M \omega,0) \chi\right]_{G^{2\ell+1}} \nonumber \\
    &+ \gamma^{\rm BH, non-universal}_{\ell m}~\,.
\end{align}
Replacing $M$ by $\mathcal{E}$ and $\chi$ by ${\cal J}$ in the first two terms
then gives the universal part of the anomalous dimension 
for a generic system, 
reproducing Eq.~\eqref{eq: universal ano}.

The fact that the anomalous dimension of the BHs is given by the BHPT renormalized angular momentum (reviewed in Supplemental Material) provides us with a detailed understanding of $\gamma^{\rm univ.}$. For instance, for 
general $\ell$, its low-orders perturbative expansion is given by 
\begin{widetext}
    \begin{equation}
    \gamma_{\ell m}^{\rm univ.} =
   -\frac{2 \left(15 \ell ^2 (\ell +1)^2+13 \ell  (\ell +1)+24\right)}{(2
   \ell +1) \ell  (\ell +1)  (4 \ell  (\ell +1)-3)} \epsilon^2
   +\frac{8 m \chi  \left(5 \ell ^3 (\ell +1)^3-\ell ^2 (\ell +1)^2+18 \ell  (\ell
   +1)+108\right)}{(2\ell+1)\ell ^2 (\ell +1)^2 (\ell  (\ell +1)-2) (4 \ell  (\ell +1)-3)}
   \epsilon^3 
   +  {\cal O}(\epsilon^4)\,,
\end{equation}
\end{widetext}
with $\epsilon = G \mathcal{E} \omega$, where the first term agrees with the well known tail prefactor \cite{Blanchet:1987wq, Almeida:2021jyt}. The explicit form through ${\cal O}(\epsilon^6)$ is given in Supplemental Material.

In particular, the quadrupolar ($\ell=2$)  universal anomalous dimensions are given by
\begin{align}
\label{eq:gamma2pert}
\gamma_{2 m}^{\rm univ.}&= - \frac{214}{105} \epsilon^2   + \frac{2 m {\cal J}}{3} \epsilon^3  - \frac{3390466}{1157625} \epsilon^4 
 + \frac{381863 m {\cal J}}{99225} \epsilon^5\,,
\end{align}
and the octupolar ($\ell=3$) one is
\begin{align}
\label{eq:gamma23ert}
 \gamma_{3m}^{\rm univ.}&=-\frac{26}{21}\epsilon^2+\frac{7m{\cal J}}{3}\epsilon^3-\frac{21842}{33957}\epsilon^4+\frac{286631 m{\cal J}}{935550}\epsilon^5\nonumber\\
& -\frac{381415329076}{481821815475}\epsilon^6+ \frac{96516668989 m {\cal J} }{136150591500}\epsilon^7 \,.
\end{align}
The first three terms of $\gamma_{2m}$ and the first term in $\gamma_{3m}$ agree with the know results~\cite{Blanchet:1997jj,Goldberger:2009qd, Trestini:2023wwg, Edison:2023qvg, Fucito:2024wlg,Almeida:2021jyt}, while the rest are new results.
Our formalism thus explicitly confirms the anomalous dimensions for electric and magnetic multipoles is the same through $\mathcal{O}(JG^{2\ell+1})$,
confirming earlier leading-order results by 
\cite{Fucito:2024wlg}.
This settles the tension 
in the literature between 
\cite{Almeida:2021jyt} and 
\cite{Fucito:2024wlg}. 
Note,
however, that while the universal magnetic and electric anomalous dimensions are the same, 
the electric and magnetic phase
shifts $\delta^{E/B}_{\ell m}$
are different, but their difference cancels in Eq.~\eqref{eq:main_anom}. 

In the eikonal limit, i.e. $G \mathcal{E} \omega \gg 1, \ell \gg 1$ but with $G \mathcal{E} \omega /\ell$ fixed, we are able to obtain the result 
\begin{align}
        \gamma_{\ell m}^{\rm univ.} \Big|_{\rm eik.} & = \ell(-1 + {}_3 F_{2}\left[ - \tfrac{1}{2} , \tfrac{1}{6} , \tfrac{5}{6}; \tfrac{1}{2},1 ; 27 x^2\right]) \nonumber  \\    & \quad + 5 m \mathcal{J} \, x^3 \, _3F_2\left[\tfrac{7}{6},\tfrac{3}{2},\tfrac{11}{6};2,\tfrac{5}{2};27 x^2\right]\,,
\end{align}

where $x= G \mathcal{E} \omega/ \ell = G \mathcal{E}/b$, and $b=\ell/\omega$ is the impact parameter. 
In this exact formula, we observe that the result has a branch cut starting at 
the impact parameter $b = 3\sqrt{3} G \mathcal{E}$, which intriguingly coincides with the radius of the BH shadow. See \cite{Parnachev:2020zbr,Akpinar:2025huz}, where some of these functions also appeared recently.

\textit{Applications to Waveform Tail Resummation.--} 
EFT allows one to compute the binary inspiral waveforms directly
from the radiative multipoles~\cite{Goldberger:2022rqf}. 
The universal anomalous dimension in Eq.~\eqref{eq: universal ano} can then be used to resum 
the ultraviolet tails in the waveform. 
This is most conveniently done in the factorized multipolar post-Minkowskian (MPM) framework \cite{Blanchet:1985sp,Blanchet:1986dk,Damour:2007yf,Damour:2008gu,Pan:2010hz,Pompili:2023tna}. 
The mode decomposition for the complex linear combination of the GW polarizations $h(t)\equiv h_+(t) - i h_\times(t)$ in terms of the spin-weight $s=-2$ spherical harmonics is
\begin{equation}
    h\left(t; \theta, \phi\right)=\sum_{\ell \geq 2} \sum_{|m| \leq \ell}{}_{-2} Y_{\ell m}\left(\theta, \phi\right) h_{\ell m}(t)\,.
\end{equation}
The mode function $h_{\ell m}(t)$ in the inspiral phase can be factorized as \cite{Damour:2007xr,Damour:2007yf,Damour:2008gu,Pan:2010hz,Pompili:2023tna}
\begin{equation}
    h_{\ell m}=h_{\ell m}^{\mathrm{N}} \hat{S}_{\mathrm{eff}} T_{\ell m} \tilde{h}_{\ell m}  ~,\label{eq: factorization}
\end{equation}
where $h_{\ell m}^{\rm N}$ is the Newtonian multipole, $\hat{S}_{\rm eff}$ the dimensionless effective source term given by either the Effective-One-Body energy $E_{\rm eff}$ \cite{Buonanno:1998gg,Buonanno:2000ef} or the orbital angular momentum $p_\phi$,  $T_{\ell m}$ is a tail resummation factor, and $\tilde{h}_{\ell m}$ is the remainder, often further decomposed in amplitude and phase as $\tilde{h}_{\ell m} = (\rho_{\ell m})^\ell e^{i \delta_{\ell m}}$. 
In this letter, we focus on improving the tail resummation $T_{\ell m}$. Physically, the tail effects capture the amplitude and the phase deflection from the wave propagation in the asymptotic background geometry.
We find it convenient to further decompose the tail part as 
\begin{equation}
    T_{\ell m} = \mathcal{S}_{\ell m} e^{i \delta^{\rm tail}_{\ell m}} ~.\label{eq: tail factor}
\end{equation}
We will refer to the amplitude $\mathcal{S}_{\ell m}$ as the Sommerfeld enhancement factor by analogy with the Coulombic scattering. 
Damour and Nagar proposed the following tail factors \cite{Damour:2007yf}
\begin{align}
    & \mathcal{S}_{\ell m} = \frac{\left|\Gamma(\ell + 1 -2 i G \mathcal{E} \omega)\right|}{\Gamma(\ell + 1)}e^{\pi G \mathcal{E} \omega}\,, \label{eq:SDN}\\
    &  \delta_{\ell m}^{{\rm tail}} = \frac12{\rm Arg}\Big[\tfrac{\Gamma(\ell + 1 - 2 i G \mathcal{E} \omega)}{\Gamma(\ell + 1 +2 i G \mathcal{E} \omega)}\Big] \!+\! (2 G \mathcal{E} \omega) \log(2 \omega r_{\rm orb}) ~,\label{eq:deltaDN}
\end{align}
which resum an infinite number of
leading (infrared) logarithms of the form $\omega^n \log^n\omega$, and associated finite parts in the Sommerfeld factor. Indeed, this form was inspired by considering wave propagation and re-scattering against the Newtonian $1/r$ potential of the binary \cite{Asada:1997zu} in the far zone, and Eqs.~\eqref{eq:SDN}-~\eqref{eq:deltaDN} correspond to the Sommerfeld factor and phase-shift for Coulombic scattering.

We are now in the position to improve upon \eqref{eq:SDN}-~\eqref{eq:deltaDN} by proposing a formula that resumms
both the infrared tails and the universal ultraviolet tails:
    \begin{align}
\mathcal{S}_{\ell m}  =& \left|\frac{\Gamma(\hat{\nu} + 1 - 2 i G \mathcal{E} \omega)}{\Gamma(\hat{\nu} + 1)}\right| e^{\pi G \mathcal{E} \omega} (r_{\rm orb}\omega)^{\hat{\nu}-\ell} ~,\label{eq: Sommerfield} \\
\delta_{\ell m}^{\rm tail}=& \frac12{\rm Arg}\Big[\tfrac{\Gamma(\hat{\nu} + 1 - 2 i G\mathcal{E} \omega)}{\Gamma(\hat{\nu} + 1 + 2 i G\mathcal{E} \omega)}\Big]+ (2 G \mathcal{E} \omega) \log(2 \omega r_{\rm orb}) \nonumber \\
&+ \frac{\ell -\hat{\nu}}{2}\pi\,,\label{eq: tail phase}
\end{align}
where the universal anomalous dimension given by Eq.~\eqref{eq: universal ano} enters as
\begin{equation}
    \hat \nu(\omega) = \ell + \gamma_{\ell m}^{\rm univ.}(\omega) \,.
\end{equation}

The factor of $(r_{\rm orb}\omega)^{\hat{\nu}-\ell}$ in the amplitude and the one proportional to $\pi$ in the phase are a direct consequence of running the RG evolution of the multipoles down to the orbital scale $\mu = 1/r_{\rm orb}$. These factors resum all universal sub-leading logarithms of the form $\omega^{n+k} \log^n \omega$ with $k>0$ corresponding to dissipative tails. The rest of the dependence on $\hat \nu$ is a proposal inspired by the test particle limit \cite{Fucito:2024wlg}, and resums additional finite terms.

This formula can be interpreted as follows: the universal tail contributions to the binary waveform are captured by the free-wave propagation in the linearized-in-spin Kerr background (i.e., the Schwarzschild–Lense–Thirring metric) sourced by the binary. The universality arises from the fact that this background is the universal part of the asymptotic metric of all compact gravitating sources. 

For quasicircular orbits $r_{\rm orb}\omega =v_{\Omega}/v_\Omega^0$, 
where $v_\Omega \equiv (G M \Omega)^{1/3}$, with $\Omega$ the orbital velocity and $M=m_1+m_2$ the total static mass of the binary system; and $v_{\Omega}^0$ is a reference velocity (see Supplemental Material). 
For gravitational waves sourced by the binary, $\omega = m \Omega$.  In this regime, we have verified our proposed resumation using the state-of-the-art PN waveform up to 4PN \cite{Faye:2014fra,Blanchet:2023sbv}, where we find that both logarithmic and $\pi$-dependent terms are resummed. 
Of course, our formula also predicts an infinite number of universal logarithms in the waveform at higher PN orders. 
We record these checks and some of these predictions in Supplemental Material. 

The formula in Eqs.~\eqref{eq: Sommerfield}-\eqref{eq: tail phase}, with anomalous dimension given in Eq.~\eqref{eq:main_anom}, does not resum all logarithms starting at 4PN order, because they contain the effects of tails-of-memory \cite{Trestini:2023wwg}, which are not universal. These depend on the intrinsic and spin-induced multipole moments of the system, and hence they cannot be simply extracted by studying the case of BHs.

\textit{Conclusions.--}
In this letter, we present the universal anomalous dimension
of the gravitational multipole 
moments of a gravitating system in general relativity. 
Using unitarity and analyticity, we derive a formula relating the anomalous dimensions of  multipole moments to the scattering phase shift of GW by the system. 
When applied to BHs, the formula identifies the multipoles anomalous dimension with the renormalized angular momentum of BHPT.
Thanks to the universality 
of the EFT action, we were able
extract the part of the BH anomalous dimension
which is universal to all compact gravitating objects 
regardless of their nature.
This conceptual advance motivates us to propose a new factorization formula for the gravitational waveform that resums all universal tails.

Our analysis provides
yet another illustration 
that EFT is a powerful tool that allows for a consistent interpretation
of the low-frequency limit of the near/far-zone expansion of GW sources.
This adds to recent progress with the definition and
extraction
of the tidal effects of a black hole
from the scattering amplitudes~\cite{Ivanov:2022qqt,Saketh:2023bul,Ivanov:2024sds,Caron-Huot:2025tlq},
which allowed one to resolve
the 
tension in the literature 
on the dynamical Love numbers
of BHs~\cite{Chakrabarti:2013lua,Charalambous:2021mea,Poisson:2020vap,Poisson:2021yau}.

The results of this letter are however limited to the universal tails. 
There are non-universal tail effects in the waveform that have not been addressed with our formula. For example, the tails-of-memory appear at 4PN order \cite{Trestini:2023wwg}, which could be beyond the description of the anomalous dimension of multipoles. This may require a new framework to deal with the worldline EFT by considering the operator algebra of multipole moments. Furthermore, various finite-size effects, which one might call tails-of-tides, enter the description beyond the orders considered here.

Going forward, it will be important to rigorously prove the factorization formulae \eqref{eq: factorization}, \eqref{eq: tail phase}, and their possible generalizations. 
Additionally, our Eq.~\eqref{eq:main_anom} strongly 
motivates the computation
of the Raman scattering of GW off
the binary, including the non-universal near-zone effects which capture the tidal deformation of the binary.
We leave these and other 
exciting research directions, 
such as the application of 
the tail-resummed waveforms to 
GW data, for future 
exploration.

\textit{Acknowledgments.}
We thank Yilber Fabian Bautista, Chia-Hsien Shen and Davide Usseglio for insightful discussions; as well as Donato Bini, Miguel Correia, Thibault Damour, Giulia Isabella and Radu Roiban for useful comments on the draft; and specially Alessandro Nagar for discussions and for providing us with computer files with the state-of-the-art PN waveforms, and their MPM resumation for comparison. YZL is supported in part by the US National Science Foundation under Grant No. PHY- 2209997, and in part by Simons Foundation grant No. 917464.

\bibliography{short.bib}


\newpage 

\pagebreak
\widetext
\begin{center}
\textbf{\large Supplemental Material}
\end{center}
\setcounter{equation}{0}
\setcounter{figure}{0}
\setcounter{table}{0}
\setcounter{page}{1}
\makeatletter
\renewcommand{\theequation}{S\arabic{equation}}
\renewcommand{\thefigure}{S\arabic{figure}}
\renewcommand{\bibnumfmt}[1]{[S#1]}
\renewcommand{\citenumfont}[1]{S#1}

\section{Definition and Computation of the Renormalized Angular Momentum}
\label{app: renormalized angular momentum}

In this appendix, we provide more details on the definition and computation of the renormalized angular momentum $\nu$. Mathematically, it is recognized as the characteristic exponent (or Floquet exponent \cite{Castro:2013kea,Castro:2013lba,Bonelli:2021uvf,Bonelli:2022ten,Bautista:2023sdf,Nasipak:2024icb}), which is derived from the Teukolsky equation. Currently, there are three methods for computing this parameter: the MST recursion relation \cite{Mano:1996gn,Mano:1996mf,Mano:1996vt,Sasaki:2003xr}, the Matone relation in terms of the Nekrasov-Shatashvili (NS) function \cite{Bonelli:2021uvf,Bonelli:2022ten,Bautista:2023sdf}, and the Monodromy matrix method \cite{Castro:2013kea,Castro:2013lba,Nasipak:2024icb}.

In the MST method, the ``renormalized" angular momentum $\nu$ is solved by the three term recurrence relation
\begin{equation}
    \alpha_n^{\nu} a_{n+1}^{\nu}+\beta_n^{\nu} a_n^{\nu}+\gamma_n^{\nu }a_{n-1}^{\nu}=0 ~,
\end{equation}
where the coefficients $\alpha_n^{\nu}, \beta_n^{\nu}$ and $\gamma_n^{\nu}$ are
\begin{equation}
\begin{aligned}
    \alpha_n^{\nu} & = \frac{i \epsilon \kappa(n+\nu+1+s+i \epsilon)(n+\nu+1+s-i \epsilon)(n+\nu+1+i \tau)}{(n+\nu+1)(2 n+2 \nu+3)} ~, \\
    \beta_n^{\nu} & =-{ }_s \lambda_{\ell}^m-s(s+1)+(n+\nu)(n+\nu+1)+\epsilon^2+\epsilon(\epsilon-m \chi)+\frac{\epsilon(\epsilon-m \chi)\left(s^2+\epsilon^2\right)}{(n+\nu)(n+\nu+1)} ~, \\
    \gamma_n^{\nu} & =-\frac{i \epsilon \kappa(n+\nu-s+i \epsilon)(n+\nu-s-i \epsilon)(n+\nu-i \tau)}{(n+\nu)(2 n+2 \nu-1)} ~,
\end{aligned}
\end{equation}
with the condition that the series $\sum_{-\infty}^\infty \alpha_n^{\nu}$ should converge both at $+\infty$ and $-\infty$.
In the above expression, the PM expansion parameter $\epsilon \equiv 2 G M \omega$, the spin-weight $s$, the dimensionless spin $\chi \equiv S/(G M^2)$, and extremality parameter $\kappa=\sqrt{1-\chi^2},\tau=(\epsilon-m \chi) / \kappa$. 

The second approach makes use of the Matone relation in the Nekrasov-Shatashvili (NS) function
\begin{equation}
    u=\frac{1}{4}-a^2+L \partial_L F\left(m_1, m_2, m_3, a, L\right)
\end{equation}
where
\begin{equation}
    \begin{aligned}
& m_1=i \frac{m \chi-\epsilon}{\kappa}, \quad m_2=-s-i \epsilon, \quad m_3=i \epsilon-s, \quad L=-2 i \epsilon \kappa, \\
& u=-{}_s\lambda_\ell^m -s(s+1)+\epsilon(i s \kappa-m \chi)+\epsilon^2(2+\kappa)\,.
\end{aligned}
\end{equation}
$a$ gives the ``renormalized'' angular momentum $a = -1/2 - \nu$ \cite{Bautista:2023sdf}. 
In the language of the four-dimensional $\mathcal{N}=2$ supersymmetric gauge theories, $m_{1,2,3}$ are the masses for the supersymmetric (hyper)multiplets, $L$ the instanton counting parameter and $a$ is the Cartan vacuum expectation value in the Coulomb branch. Mathematically, $a$ is also known as the the quantum A-period of the confluent Heun equation. $F$ is the NS function, which is essentially the instanton part of the NS free energy \cite{Bonelli:2021uvf,Bautista:2023sdf}. This approach provides us with formal understanding of the structure of $\nu$ even in the high frequency limit
\begin{equation}
    \nu \simeq -2i G M \omega \quad  {\rm as}  \quad G M \omega \gg 1 ~.
\end{equation}

The third approach is closely related to the second one, and it provides a mathematical interpretation of the ``renromalized" angular momentum by studying the monodromy matrix around the irregular singular points of the confluent Heun equations. Formally, the monodromy matrix around the irregular singular points at infinity takes the form
\begin{equation}
    M_{\infty} = 
    \begin{pmatrix}
        e^{2 \pi i \nu_{\infty}} & 0\\
        0 & e^{-2 \pi i \nu_{\infty}}  
    \end{pmatrix}
    ~,
\end{equation}
where $\nu_{\infty}$ is the characteristic exponent that can be evaluated by solving stokes parameters. In Ref.~\cite{Nasipak:2024icb}, the author has shown that the ``renormalized'' angular momentum is precisely the characteristic exponent $\nu_{\infty}$.

In any of the above three methods, one can solve the ``renormalized" angular momentum of BHs perturbatively, i.e. $\nu = \ell + \nu_n (G M \omega)^n, n =2,3 \cdots$. Here, we explicitly show the generic $\ell$ expressions for $\nu_n,n=2,3,4,5,6,7$ through linear order in spin, where the first three terms agrees with Ref.~\cite{Sasaki:2003xr,Fucito:2024wlg} and the even-$n$ ones agree with Ref.~\cite{Bini:2013rfa}
\begin{align}
    \nu_2 & =  -\frac{2 \left(15 \lambda ^2+13 \lambda +24\right)}{(2
   \ell +1) \ell  (\ell +1)  (4 \ell  (\ell +1)-3)} ~, \\
    \nu_3 & = \frac{8 m \chi  \left(5 \lambda ^3-\lambda ^2+18 \lambda +108\right)}{(\ell -1) \ell ^2 (\ell +1)^2 (\ell +2) (2 \ell -1) (2 \ell +1) (2 \ell
   +3)} \label{eq:general ell nu3} ~, \\
   \nu_4 & = \frac{2(-18480 \lambda ^8+61320 \lambda ^7-2415 \lambda ^6+85775 \lambda ^5+123233 \lambda ^4+51522 \lambda ^3-953424 \lambda ^2+102816 \lambda +51840)}{(\ell -1) \ell ^3 (\ell +1)^3 (\ell +2)
   (2 \ell -3) (2 \ell -1)^3 (2 \ell +1)^3 (2 \ell +3)^3 (2 \ell +5)}  \label{eq:general ell nu4} ~, \\
   \nu_5 & = \frac{48 m \chi(3696 \lambda ^9-13944 \lambda ^8+18347 \lambda ^7-22136 \lambda ^6-42625 \lambda ^5-145050 \lambda ^4-650274 \lambda ^3+1450620 \lambda ^2-125064 \lambda -77760)}{(\ell
   -1)^2 \ell ^4 (\ell +1)^4 (\ell +2)^2 (2 \ell -3) (2 \ell -1)^3 (2 \ell +1)^3 (2 \ell
   +3)^3 (2 \ell +5)} \label{eq:general ell nu5} ~, \\
   \nu_6 & = -\frac{4}{(\ell -1)^2 \ell ^5 (\ell +1)^5 (\ell
   +2)^2 (2 \ell -5) (2 \ell -3)^2 (2 \ell -1)^5 (2 \ell +1)^5 (2 \ell +3)^5 (2 \ell
   +5)^2 (2 \ell +7)} \Bigg[104552448 \lambda ^{15} \nonumber \\
   & \quad -1671301632 \lambda ^{14}+8204035840 \lambda ^{13}-15243669056 \lambda ^{12}+13732238520 \lambda ^{11}-12944646946 \lambda ^{10}-13002690896 \lambda
   ^9 \nonumber \\
   & \quad -24635974293 \lambda ^8+887441317 \lambda ^7+30247168320 \lambda ^6+680072616180 \lambda ^5-1013061463920 \lambda ^4+111802065696 \lambda ^3 \nonumber \\
   & \quad +82127701440 \lambda
   ^2-12975033600 \lambda -3919104000 \Bigg] \label{eq:general ell nu6} ~, \\
   \nu_7 & = \frac{48 m \chi }{(\ell -2) (\ell -1)^3 \ell ^6 (\ell +1)^6 (\ell +2)^3 (\ell +3) (2 \ell -5) (2 \ell -3)^2 (2 \ell -1)^5 (2 \ell +1)^5 (2 \ell +3)^5 (2 \ell +5)^2 (2 \ell +7)} \nonumber \\
   & \times \Bigg[74680320 \lambda ^{17}-1644797440 \lambda ^{16}+13439345920 \lambda ^{15}-53051339968 \lambda ^{14}+115693152168 \lambda ^{13}-153954147622 \lambda ^{12} \nonumber \\
   & \quad +104796913232 \lambda
   ^{11}-46104555329 \lambda ^{10}+51933011989 \lambda ^9+352999107060 \lambda ^8-571463718576 \lambda ^7 \nonumber \\
   & \quad +11287693868616 \lambda ^6  -33483100996872 \lambda ^5+28193417777664
   \lambda ^4-1752702484032 \lambda ^3-2381917337280 \lambda ^2 \nonumber \\
   & \quad +293009011200 \lambda +105815808000\Bigg] ~,
\end{align}
where we have introduced $\lambda \equiv \ell (\ell+1)$.
Note that these generic-$\ell$ expressions are only valid for integer $\ell > \ell^*$ where $\ell^*$ is the location of the largest pole in the denominator, of the form $1/(\ell - \ell^*)$. For instance, the ${\cal O}(G^6)$ correction to $\nu$ in Eq.~\eqref{eq:general ell nu6} has a pole at $\ell = 5/2$, so the formula is not to be trusted for $\ell=2$ starting at this order.
The breakdown of the generic-$\ell$ low-frequency expansion of $\nu$ is related to the existence of (running) dynamical tides starting at this order \cite{Ivanov:2024sds}.

\section{Post-Newtonian Checks of Waveform Tail Resummation}
\label{waveform}

In this appendix, we show that the improved tail factors \eqref{eq: tail factor}, \eqref{eq: Sommerfield}, and \eqref{eq: tail phase} can indeed improve the PN-expanded waveform by resumming the tail logarithms and their finite associates by comparing to the know results up to the 4PN order for the $(\ell,m)=(2,2)$ waveform \cite{Blanchet:2023sbv}, and 3.5PN for the $(3,1)$ and $(3,3)$ waveform \cite{Faye:2014fra}.

We now set the convention to align with \cite{Faye:2014fra,Blanchet:2023sbv}, where they factorized a phase factor
\be
h_{\ell m}=\frac{8G M \eta x}{R}\sqrt{\frac{\pi}{5}}H_{\ell m}e^{-im\psi}\,.
\ee
Here, $M \equiv m_1+ m _2$ is the total mass of the binary, $\eta\equiv m_1 m_2/ (m_1 + m_2)^2$ is the symmetric mass ratio, $x$ is the PN-expansion parameter $x=v^2=(GM\Omega)^{2/3}$, $R$ is the radiative radial coordinate, and $\Omega$ is the measurable GW half-frequency. The phase factor $\psi$ is chosen by hand to be \cite{Faye:2014fra}
\be
\psi=\varphi-2G\mathcal{E}\omega \log\frac{\omega}{\omega_0}\,,\quad \log(4 \omega_0 b)=\frac{11}{12}-\gamma_E\,,
\ee
where $b$ is a reference time scale. Considering the radiative coordinates and harmonic coordinates $T_R=t_r-2G\mathcal{E}\log(r/b)$ and the logarithmic separation $\log(\omega r)=\log(\omega_0 b)+\log(\omega/\omega_0)+\log(r/b)$, our factorization formula for $H_{\ell m}$ gives
\be
H_{\ell m}=H_{\ell m}^N\hat{S}_{\rm eff}T_{\ell m}e^{2iG\mathcal{E}\omega\log\big(\frac{\omega_0 b}{\omega r_{\rm orb}}\big)} \tilde{H}_{\ell m} \,,\label{eq: resum H}
\ee
where we choose the reference velocity to be $v_\Omega^0=\omega_0 b$.

Ref.~\cite{Blanchet:2023sbv} obtained $H_{22}$ up to 4PN order, which is given by 
\begin{align}
& H_{22}=1+\left[-\frac{107}{42}+\frac{55  }{42} \eta\right] x+{\color{red}{2\pi}} x^{\frac{3}{2}}+\left[-\frac{2173}{1512}-\frac{1069  }{216}\eta+\frac{2047 }{1512}\eta ^2\right] x^2+\left[-{\color{red}\frac{107 \pi }{21}}+\left({\color{red}\frac{34 \pi }{21}}-24 i\right) \eta \right] x^{\frac{5}{2}}\nonumber\\
&+\left[\left({\color{violet}{-\frac{428}{105} \log (16 x)+\frac{2 \pi ^2}{3}-\frac{856 \gamma_E }{105}+\frac{428 i \pi }{105}}}+\frac{27027409}{646800}\right)+\left(\frac{41 \pi ^2}{96}-\frac{278185}{33264}\right) \eta-\frac{20261 }{2772}\eta ^2+\frac{114635 }{99792}\eta ^3\right]x^3 \nonumber\\
&+\left[ {\color{red}-\frac{2173 \pi }{756}}+\left({\color{red}-\frac{2495 \pi }{378}}+\frac{14333 i}{162}\right) \eta+\left({\color{red}\frac{40 \pi }{27}}-\frac{4066 i}{945} \right) \eta ^2\right] x^{\frac{7}{2}}+ \Bigg[\Bigg({\color{violet}{\frac{22898 \log (16 x)}{2205}+\frac{45796 \gamma_E }{2205}-\frac{22898 i \pi }{2205}-\frac{107 \pi ^2}{63}}}\nonumber\\
&-\frac{846557506853}{12713500800}\Bigg)+\left(\frac{7642}{441} \log (16 x)-\frac{336005827477}{4237833600}+\frac{15284 \gamma_E }{441}-\frac{219314 i \pi }{2205}-\frac{9755 \pi ^2}{32256}\right)\eta\nonumber\\
&+\left(\frac{256450291}{7413120}-\frac{1025 \pi ^2}{1008}\right) \eta ^2 -\frac{81579187 }{15567552}\eta ^3+\frac{26251249 \eta^4}{31135104}\Bigg]x^4+\mathcal{O}(x^{\frac{9}{2}})\,,
\end{align}
where red terms are those that the resummation formula of the {\color{red} IR} tails \cite{Damour:2007xr,Damour:2007yf,Damour:2008gu} can improve, while violet terms are our resummation formula \eqref{eq: resum H} can further improve by also resumming {\color{violet} UV} tails. In particular, violet terms are fully resummed, while the red terms are improved as their transcendental weights are lowered (from $T[\pi]=1$ to $T[\text{rational}]=0$). Note that the logarithm and other transcendental terms at the $x^4 \eta$ order are not resummed because they include the tails-of-memory effects \cite{Trestini:2023wwg} that are beyond the scope of tail resummation from multipole anomalous dimensions. Nevertheless, the numbers for other rational terms are simplified by our resummation.

To explicitly show this, we use the energy and angular momentum of the binary up to 4PN from \cite{Bernard:2017ktp}
\begin{align}
&\frac{\mathcal{E}}{M}=1-\frac{\eta x}{2}+\left[\frac{3 \eta }{8}+\frac{\eta ^2}{24}\right]x^2+\left[\frac{\eta ^3}{48}-\frac{19 \eta ^2}{16}+\frac{27 \eta }{16}\right]x^3+\left[\frac{675\eta}{128}+\left(\frac{205 \pi ^2}{192}-\frac{34445}{1152}\right) \eta ^2+\frac{155 \eta ^3}{192}+\frac{35 \eta ^4}{10368}\right]x^4\nonumber\\
&+\left[\frac{3969 \eta }{256}+ \left(-\frac{224 \log (16 x)}{15} -\frac{448 \gamma_E }{15}-\frac{9037 \pi ^2}{3072}+\frac{123671}{11520}\right)\eta^2+\left(\frac{498449}{6912}-\frac{3157 \pi ^2}{1152}\right) \eta ^3-\frac{301 \eta ^4}{3456}-\frac{77 \eta ^5}{62208}\right]x^5\,,\nonumber\\
& \frac{\sqrt{x} J}{GM\mu}=1+\left[\frac{3}{2}+\frac{\eta }{6}\right] x+\left[\frac{27}{8}-\frac{19 \eta }{8}+\frac{\eta ^2}{24}\right]x^2+\left[\frac{135}{16}+\left(\frac{41 \pi ^2}{24}-\frac{6889}{144}\right) \eta+\frac{31 \eta ^2}{24}+\frac{7 \eta ^3}{1296}\right]x^3+\Bigg[\frac{2835}{128}\nonumber\\
&+\eta  \left(-\frac{64}{3} \log (16 x)-\frac{128 \gamma_E }{3}-\frac{6455 \pi ^2}{1536}+\frac{98869}{5760}\right)+\left(\frac{356035}{3456}-\frac{2255 \pi ^2}{576}\right) \eta ^2-\frac{215 \eta ^3}{1728}-\frac{55 \eta ^4}{31104}\Bigg]x^4\,,\label{eq:EJcons}
\end{align}
where $\mu=m_1m_2/M$ is the reduced mass. 
For even $\ell+m$, the effective source term $\hat{S}_{\rm eff}$ is the effective Hamiltonian $\hat{H}_{\rm eff}$, which is related to total energy by
\be
\mathcal{E}=M \sqrt{1+2\eta \big(\hat{H}_{\rm eff}-1\big)}\,,\quad \hat{H}_{\rm eff}=\frac{H_{\rm eff}}{\eta}\,.
\ee
Dividing $H_{22}$ by $\hat{H}_{\rm eff}T_{22}e^{2iG\mathcal{E}\omega\log\big(\omega_0b/(\omega r_{\rm orb})\big)}$ which captures the tail resummation, we find
\begin{align}
& H_{22}^{N}\tilde{H}_{22} =1+\left[-\frac{43}{21}+\frac{55  }{42} \eta\right] x+\frac{7i}{3} x^{\frac{3}{2}}+\left[-\frac{536}{189}-\frac{6745  }{1512}\eta+\frac{2047 }{1512}\eta ^2\right] x^2+\left[-\frac{43 i }{9}-\frac{199i }{9} \eta\right] x^{\frac{5}{2}}\nonumber\\
&+\left[\frac{7004896}{363825}+\left(\frac{41 \pi ^2}{96}-\frac{34625}{3696}\right) \eta-\frac{227875 }{33264}\eta ^2+\frac{114635 }{99792}\eta ^3\right]x^3 +\left[ -\frac{536 i }{81}+\frac{22463i}{324}\eta-\frac{7298i}{2835} \eta^2\right] x^{\frac{7}{2}}\nonumber\\
& +\Bigg[-\frac{622302262}{99324225}+\left(\frac{464}{35} \log (16 x)+\frac{42582952999}{12713500800}+\frac{928 \gamma_E }{35}-\frac{4976 i \pi }{105}-\frac{43963 \pi ^2}{32256}\right)\eta\nonumber\\
&+\left(\frac{20365047}{640640}-\frac{1025 \pi ^2}{1008}\right) \eta ^2-\frac{76054213 }{15567552}\eta ^3+\frac{26251249 }{31135104} \eta^4 \Bigg]x^4+\mathcal{O}(x^{\frac{9}{2}})\,,
\end{align}

Similarly, we can also resum the 3.5PN tail in $H_{33}$ and $H_{31}$, which are given by 
\cite{Faye:2014fra}
\begin{align}
&H_{33}=-\frac{3}{4}i\sqrt{\frac{15}{14}}\sqrt{1-4\eta}\Bigg[\sqrt{x}+\left[-4+2\eta\right]x^{\frac{3}{2}}+\left[{\color{red} 3\pi+6i\log\Big(\frac{3}{2}\Big)}-\frac{21i}{5}\right]x^2+\left[\frac{123}{110}-\frac{1838  }{165} \eta+\frac{887 }{330} \eta ^2\right] x^{\frac{5}{2}}\nonumber\\
&+ \left[{\color{red}-12\pi-24i\log\Big(\frac{3}{2}\Big)}+\frac{84i}{5}+\left({\color{red} \frac{9\pi}{2}+9i\log\Big(\frac{3}{2}\Big)}-\frac{48103i}{1215}\right)\eta\right]x^3+\Bigg[{\color{violet}\Big(-\frac{39}{7} \log (16 x)+\frac{3 \pi ^2}{2}-\frac{78 \gamma_{E} }{7} }\nonumber\\
&{\color{violet} +6 i \pi  \left(3 \log \Big(\frac{3}{2}\Big)-\frac{41}{35}\right) }{\color{red}-18\log^2\Big(\frac{3}{2}\Big)}+\frac{19388147}{280280}\Big)+\left(\frac{41 \pi ^2}{64}-\frac{7055}{3432}\right)\eta-\frac{318841}{17160} \eta ^2+\frac{8237}{2860} \eta ^3\Bigg]x^{\frac{7}{2}}\Bigg]\,,\nonumber\\
& H_{31}=i\frac{\sqrt{1-4\eta}}{12\sqrt{14}}\Bigg[\sqrt{x}+\left[-\frac{8}{3}-\frac{2}{3}\eta\right]x^{\frac{3}{2}}+\left[-\frac{7i}{5} {\color{red}+\pi -2 i \log (2)}\right]x^2+\left[\frac{607}{198}-\frac{136  }{99}\eta-\frac{247 }{198}\eta ^2\right] x^{\frac{5}{2}}\nonumber\\
&+ \left[\left(\frac{56 i}{15}{\color{red}-\frac{8 \pi }{3}+\frac{16}{3} i \log (2)}\right)+ \left(-\frac{i}{15}{\color{red}-\frac{7 \pi }{6}+\frac{7}{3} i \log (2)}\right)\eta \right]x^3+\Bigg[{\color{violet}-\frac{13\log (16 x)}{21} +\frac{\pi ^2}{6}-\frac{26 \gamma_{E} }{21}-\frac{82 i \pi }{105}}\nonumber\\
&{\color{red} -2 \log ^2(2) -2 i \pi  \log (2)}-\frac{164 \log (2)}{105}+\frac{10753397}{1513512}+\left(\frac{41 \pi ^2}{64}-\frac{1738843}{154440}\right) \eta+\frac{327059}{30888} \eta ^2-\frac{17525}{15444}\eta ^3\Bigg]x^{\frac{7}{2}} \Bigg]\,.
\end{align}
After our resummation, we find
\begin{align}
& H_{33}^N \tilde{H}_{33}=-\frac{3}{4}i\sqrt{\frac{15}{14}}\sqrt{1-4\eta}\Bigg[\sqrt{x}+\left[-\frac{7}{2}+2\eta\right]x^{\frac{3}{2}}+\frac{13i}{10}x^2+\left[-\frac{443}{440}-\frac{3401}{330}  \eta +\frac{887}{330} \eta ^2\right]x^{\frac{5}{2}}\nonumber\\
&+ \left[-\frac{ 91 i}{20}-\frac{152317 i }{4860} \eta\right]x^3+\left[\frac{23294919}{560560}-\frac{78}{7} \log \Big(\frac{3}{2}\Big)+\left(\frac{41 \pi ^2}{64}-\frac{17161}{2860}\right) \eta-\frac{27409}{1560} \eta ^2+\frac{8237 }{2860}\eta ^3\right] x^{\frac{7}{2}}\Bigg]\,,\nonumber\\
& H_{31}^N \tilde{H}_{31}=i\frac{\sqrt{1-4\eta}}{12\sqrt{14}}\Bigg[\sqrt{x}+\left[-\frac{13}{6}-\frac{2}{3}\eta\right]x^{\frac{3}{2}}+\frac{13 i x^2}{30}+\left[\frac{1273}{792}-\frac{371 }{198} \eta -\frac{247}{198}  \eta ^2 \right]x^{\frac{5}{3}}\nonumber\\
&+\left[-\frac{169 i}{180} -\frac{397 i  }{180} \eta\right]x^3+\left[\frac{61487333}{15135120}+\frac{26}{21}  \log (2)+\left(\frac{41 \pi ^2}{64}-\frac{788399}{77220}\right) \eta+\frac{311225 }{30888}\eta ^2-\frac{17525 }{15444}\eta ^3\right]x^{\frac{7}{2}}\Bigg]\,.
\end{align}

Moreover, the universal anomalous dimensions enable us to predict the corresponding logarithmic structures in the waveform at higher post-Newtonian (PN) orders, which is beyond the current reach of PN calculations. For instance, in the probe limit, where tail-of-memory effects can be neglected, we predict:
\begin{align}
& H_{22}^{{\rm univ}\, {\rm log}}=\left(-\frac{428 }{105}x^3+\frac{22898}{2205} x^4-\frac{ 856 \pi }{105} x^{9/2}+\frac{45796 \pi }{2205}x^{11/2}+\cdots\right)\log x\,,\nonumber\\
& H_{33}^{{\rm univ}\, {\rm log}}=\left(-\frac{39}{7}  x^{7/2}+\frac{156}{7} x^{9/2}-\frac{117}{35} \left(-7 i+5 \pi +10 i \log \left(\frac{3}{2}\right)\right)x^5 +\frac{468}{35} \left(-7 i+5 \pi +10 i \log \left(\frac{3}{2}\right)\right) x^6+\cdots\right)\log x\,,\nonumber\\
& H_{31}^{{\rm univ}\, {\rm log}}=\left(-\frac{13}{21}  x^{7/2}+\frac{104 }{63}x^{9/2}+\left(\frac{13 i}{15}-\frac{13 \pi }{21}+\frac{26}{21} i \log (2)\right)x^5+\left(-\frac{104 i}{45}+\frac{104 \pi }{63}-\frac{208}{63} i \log (2)\right)x^6 + \cdots\right)\log x\,.
\end{align}

\end{document}